\documentclass[reprint,showpacs,amsmath,amssymb,pra]{revtex4-1}
\usepackage{graphicx}
\usepackage{dcolumn}
\usepackage{bm}

\begin{document}

\title{Controlled absorption and all-optical diode action due to collisions of self-induced transparency solitons}

\author{Denis V. Novitsky}
\email{dvnovitsky@tut.by} \affiliation{B. I. Stepanov Institute of
Physics, National Academy of Sciences of Belarus, Nezavisimosti
Avenue 68, BY-220072 Minsk, Belarus}

\date{\today}

\begin{abstract}
We study inelastic collisions of counter-propagating self-induced
transparency solitons in a homogeneously broadened two-level medium.
The energy of the pulse can be almost totally absorbed in the medium
due to asymmetric collision with a properly chosen control pulse.
The medium state thus prepared demonstrates the property of an
all-optical diode which transmits pulses from one direction and
blocks from another. The saturation process of a controlled
absorption effect, local-field correction influence, and the
parameter ranges for the diode action are studied as well.
\end{abstract}


\maketitle

\section{Introduction}

The problem of creation of all-optical diodes has been actively
studied during recent years. Such systems, which transmit
electromagnetic radiation only in one direction and block it in
another, are usually based on asymmetrically constructed waveguides
\cite{Gallo} or nonlinear photonic structures (see, for example,
\cite{Zhukovsky, Cuicui, Chunhua, Cai}). The use of left-handed
materials \cite{Feise} and asymmetrically absorbing systems
\cite{Philip} have been studied, among other possibilities. In other
words, the all-optical diode action is commonly a consequence of the
proper ordering of optical elements on microscopic scale (of the
order of a light wavelength).

In this paper we propose a fundamentally different scheme of
all-optical diode action. It is entirely based on nonlinear dynamics
of the self-induced transparency (SIT) solitons in a homogeneously
broadened, dense two-level medium. SIT is a well-known effect
\cite{McCall, Poluektov}, resulting in formation of stationary
pulses (solitons) which propagate without change in their form. Our
approach deals with collisions of counter-propagating SIT solitons.
The effects of such collisions were partially studied in our
previous work \cite{Novit2}. It was shown that under certain
conditions, one of the colliding solitons can be almost entirely
absorbed by the medium. We call this phenomenon \textit{the
controlled absorption} of the SIT soliton.

A large density of the two-level medium allows one to obtain the
resulting effects on shorter distances. It is also useful for
reducing the scope of the numerical calculations. The local-field
effects need to be taken into account in this case. In Ref.
\cite{Novit1} it was shown that they do not significantly affect the
dynamics of a pulse with duration of femtosecond order. However, the
local-field correction is taken into account here for generality;
moreover, it is possible that it can somehow influence the dynamics
on relatively long distances, as in our consideration. Therefore, we
discuss this problem further as well.

The paper consists of the following sections. In Sec. \ref{abs} we
consider the effect of controlled absorption of the SIT soliton in
some detail. Section \ref{diode} is devoted to an explanation of our
mechanism of all-optical diode action. In the Sec. \ref{time} we
discuss the influence of pulse launch time, relaxation time, and
medium density (as well as local field) on the demonstration of our
effects. Finally, in Sec. \ref{concl} we make some concluding
remarks.

\section{\label{abs}Controlled absorption}

Propagation of light pulse in the dense two-level medium is governed
by the system of semiclassical Maxwell-Bloch equations for
population difference $W$, microscopic polarization $R$, and
dimensionless Rabi frequency
$\Omega'=\Omega/\omega=(\mu/\hbar\omega)E$ (i.e., the normalized
electric field amplitude) \cite{Bowd93, Cren96}:
\begin{eqnarray}
\frac{dR}{d\tau}&=& i \Omega' W + i R (\delta+\epsilon W) - \gamma_2 R, \label{dPdtau} \\
\frac{dW}{d\tau}&=&2 i (\Omega'^* R - R^* \Omega') -
\gamma_1 (W-1), \label{dNdtau} \\
\frac{\partial^2 \Omega'}{\partial \xi^2}&-& \frac{\partial^2
\Omega'}{\partial \tau^2}+2 i \frac{\partial \Omega'}{\partial
\xi}+2 i \frac{\partial \Omega'}{\partial
\tau} \nonumber \\
&&=3 \epsilon \left(\frac{\partial^2 R}{\partial \tau^2}-2 i
\frac{\partial R}{\partial \tau}-R\right), \label{Maxdl}
\end{eqnarray}
where $\tau=\omega t$ and $\xi=kz$ are dimensionless time and
distance, respectively; $\delta=\Delta\omega/\omega$ the normalized
detuning of the field carrier (central) frequency $\omega$ from the
atomic resonance; $\gamma_1=(\omega T_1)^{-1}$ and $\gamma_2=(\omega
T_2)^{-1}$ the rates of longitudinal and transverse relaxation,
respectively; $k=\omega/c$ the wavenumber; $c$ the light speed in
vacuum; and $\epsilon=\omega_L/ \omega = 4 \pi \mu^2 C / 3 \hbar
\omega$ is the normalized Lorentz frequency, where $C$ is the
concentration of the two-level atoms, and $\mu$ the component of
transition dipole moment parallel to the polarization vector of the
electric field. Here we assume without loss of generality that the
background dielectric permittivity of the medium is unity (two-level
atoms in vacuum). Note that in Eq. (\ref{Maxdl}) we do not use the
slowly varying envelope approximation (SVEA), while in Eq.
(\ref{dPdtau}) the term $\epsilon W$ is responsible for the
so-called local field correction (near dipole-dipole interaction) in
a dense medium. The numerical approach used in this paper to solve
Eqs. (\ref{dPdtau})--(\ref{Maxdl}) can be found in Ref.
\cite{Novit}.

We consider propagation of ultrashort pulses with Gaussian shape
$\Omega=\Omega_p\exp(-t^2/2t_p^2)$, where $t_p$ is the pulse
duration. The amplitude of the pulses is measured in the units of
the characteristic Rabi frequency $\Omega_0=\sqrt{2\pi}/2 t_p$,
which corresponds to the so-called $2 \pi$ pulse. In our
calculations the values $T_1=1$ ns and $T_2=0.1$ ns are taken so
that femtosecond pulses are in the regime of coherent interaction
with the resonant medium. We use the following parameters of
calculations which hold true throughout the paper:
$\omega_L=10^{11}$ s$^{-1}$, $\delta=0$ (exact resonance),
$\lambda=2 \pi c / \omega=0.5$ $\mu$m; the thickness of the medium
$L=1000 \lambda$. All the pulses have the same duration $t_p=50$ fs,
which means that the only parameter governing pulse dynamics is its
amplitude. One can estimate that a $2 \pi$ pulse with such duration
should have peak intensity of about $10$ MW/cm$^2$ in the case of
atomic dipole moments $\mu \sim 1$ D. The required concentration in
this case is $C \sim 10^{19}$ cm$^{-3}$. However, the density and
the peak intensity can be significantly reduced if the media with
higher dipole moments is used (for example, a collection of quantum
dots).

\begin{figure}[t!]
\includegraphics[scale=0.88, clip=]{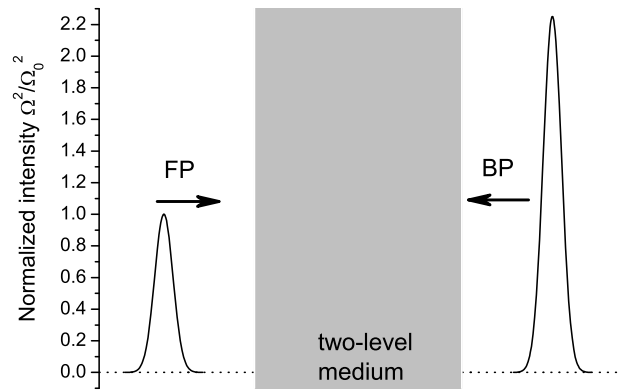}
\caption{\label{fig1} The scheme of collision of two
counter-propagating (FP and BP) pulses in a two-level medium.}
\end{figure}

We start with the scheme of two ultrashort (coherent)
counter-propagating pulses: the forward-propagating (FP) pulse with
the initial amplitude $\Omega_{p1}$ and the backward-propagating
(BP) one with $\Omega_{p2}$ (see Fig. \ref{fig1}). If their
amplitudes are large enough, both pulses form self-induced
transparency (SIT) solitons as they propagate inside the medium.
Such two counter-propagating solitons meet somewhere inside the
medium and interact. It is known that, in contrast to the pair of
co-propagating solitons, this interaction is inelastic \cite{Shaw}.
Mathematically, this means that two counter-propagating pulses do
not form a stationary solution of the equations of motion. From the
physical point of view, this fact can be explained with the simple
argumentation as follows. For the co-propagating solitons, we always
have a situation when the light energy absorbed at the leading edge
of the pulse is released at its trailing edge. This is not the case
for the counter-propagating solitons: at the point of collision, the
medium excited at the leading edge of the first pulse interacts with
the second pulse even before the trailing edge arrives. As a result,
the collision of the pulses leads to the partial absorption of their
energy.

\begin{figure}[t!]
\includegraphics[scale=0.88, clip=]{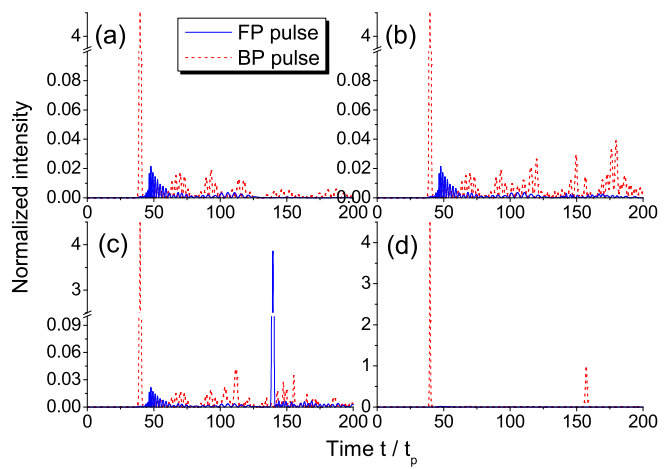}
\caption{\label{fig2} (Color online) The resulting intensities after
the collision: (a) collision of two pulses with
$\Omega_{p1}=\Omega_0$ and $\Omega_{p2}=1.5\Omega_0$; (b) the third
(probe) FP pulse $\Omega_{p3}=\Omega_0$ is added; (c) the probe FP
pulse has $\Omega_{p3}=1.5\Omega_0$; (d) the probe pulse with
$\Omega_{p3}=\Omega_0$ is BP. The first and second pulses start at
the instant $t=0$, while the third one starts at $t=100 t_p$.}
\end{figure}

It was recently shown \cite{Novit2} that, if $\Omega_{p1}=\Omega_0$
($2\pi$-pulse) and $\Omega_{p2}=1.5\Omega_0$ ($3\pi$-pulse), the FP
soliton gets entirely absorbed, while the BP one appears unperturbed
at the output. In other words, the first pulse is blocked by the
second one. For simplicity, we call this situation the controlled
absorption of light energy inside the two-level medium. This energy
leaves the medium in a long run as a result of fluorescence. We
should also emphasize that this effect occurs only for the
asymmetric collision, i.e. when the amplitudes of both pulses are
not the same. The resulting profiles of the FP and BP pulses with
initial amplitudes $\Omega_{p1}=\Omega_0$ and
$\Omega_{p2}=1.5\Omega_0$ are demonstrated in Fig. \ref{fig2}(a).
These profiles are recorded at the opposite edges of the medium:
according to Fig. \ref{fig1}, it is right edge for the FP pulse and
the left edge for the BP one. One can easily see that the FP pulse
is absent at the output (its energy is mostly absorbed), while the
BP pulse appears at the other end of the medium. Note that this
pulse (we can call it the control pulse) can be described by the
usual $2 \pi$-soliton envelope and is significantly compressed in
comparison with the initial pulse due to the transient process
\cite{Novit2}. There are also low-intensity tails connected with
radiation of absorbed light at both outputs of the medium.

\section{\label{diode}Diode action}

Now one can ask a question: what happens when the third (or probe)
pulse with the amplitude $\Omega_{p3}$ enters the medium? There are
three possible answers: (i) it will be trapped, (ii) it will pass
through the medium, and (iii) it will not only pass but also
retrieve the previously absorbed pulse from the medium. Further we
try to investigate this question for different values of
$\Omega_{p3}$. If the possibility (iii) can be realized, one can
talk about optical memory.

Let us add a probe pulse with amplitude $\Omega_{p3}$ which enters
the medium after the interval $t=100 t_p$ when there is no
propagating solitons inside the medium. There are two alternatives:
the probe pulse is FP or BP. The case of FP probe pulse with
$\Omega_{p3}=\Omega_0$ is shown in Fig. \ref{fig2}(b). It is seen
that interaction of this pulse with the medium (which stores the
energy of the previous FP pulse) prevents its appearance at the
output. Only tails have somewhat increased intensity. As opposed to
this case, the pulse $\Omega_{p3}=1.5 \Omega_0$ is happily
transmitted through the medium though with slightly lower peak
intensity than the identical BP pulse [Fig. \ref{fig2}(c)]. This
means that transmittance of the soliton depends on its intensity.
Moreover, we have an interesting deviation from the area theorem:
the pulse with initial area $2 \pi$ gets absorbed, while the one
with $3 \pi$ forms a $2 \pi$-soliton.

Now, if the probe pulse is BP and $\Omega_{p3}=\Omega_0$, then it
appears at the output [Fig. \ref{fig2}(d)], in sharp contrast with
the FP case [Fig. \ref{fig2}(b)]. In other words, the collision of
two counter-propagating pulses switches the medium into such a state
that it transmits the probe pulse incident from one side and blocks
from the other. We can call these effect \textit{the all-optical
diode action}. The reason for this effect can be understood if we
recall that the colliding pulses are not identical, i.e. the
collision is asymmetric. As a consequence, this asymmetric collision
"prepares" the asymmetric state of the medium, i.e. the diode action
entirely depends on light-matter interaction, but not on the
properly chosen geometry of the problem. Therefore, this scheme is
fundamentally different from usual proposals of all-optical diodes
based on asymmetrically constructed systems.

\begin{figure}[t!]
\includegraphics[scale=0.9, clip=]{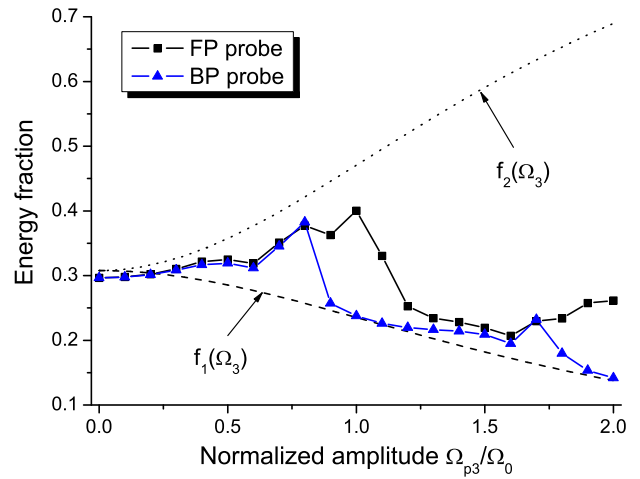}
\caption{\label{fig3} (Color online) The dependence of absorbed
energy (as a fraction of the total energy) on the amplitude
$\Omega_{p3}$ of the probe pulse. The energy is integrated over the
time period $t=200 t_p$.}
\end{figure}

\begin{figure}[t!]
\includegraphics[scale=0.88, clip=]{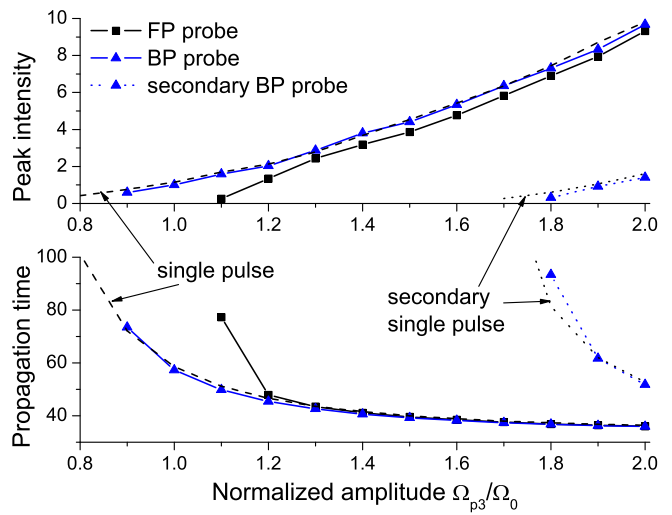}
\caption{\label{fig4} (Color online) The dependence of peak
intensity and propagation time (in units of $t_p$) on the amplitude
$\Omega_{p3}$ of the probe pulse. For comparison, the results for a
single pulse with the amplitude $\Omega_{p3}$ are shown.}
\end{figure}

To study this asymmetry in more detail and to clarify the conditions
for the diode action, we investigate the dependence of absorbed
energy on the amplitude of the probe pulse $\Omega_{p3}$ (see Fig.
\ref{fig3}). Absorbed energy is calculated as a part of total
electromagnetic energy which remains inside the medium after some
period (namely, $t=200 t_p$) from the start of the first two pulses.
The curves for both (FP and BP) cases are situated between two
asymptotes. The first one characterizes the fraction of the total
energy of all three pulses which accounts for the first FP pulse.
This asymptote is defined as the function $f_1
(\Omega_{p3})=\Omega_{p1}^2/(\Omega_{p1}^2+\Omega_{p2}^2+\Omega_{p3}^2)=1/(3.25+\tilde\Omega_{p3}^2)$,
where we use the values $\Omega_{p1}=\Omega_0$,
$\Omega_{p2}=1.5\Omega_0$ and
$\tilde\Omega_{p3}=\Omega_{p3}/\Omega_0$. It describes the absorbed
energy in the ideal case when only the energy of the first pulse is
trapped in the medium. Similarly, we introduce the second asymptote
$f_2
(\Omega_{p3})=(\Omega_{p1}^2+\Omega_{p3}^2)/(\Omega_{p1}^2+\Omega_{p2}^2+\Omega_{p3}^2)=(1+\tilde\Omega_{p3}^2)/(3.25+\tilde\Omega_{p3}^2)$,
which describes the part of the energy carried by the first and
third pulses together. It corresponds to another ideal situation
when both the first and the probe pulses are absorbed.

It is seen in Fig. \ref{fig3} that at low values of $\Omega_{p3}$,
the absorbed energy curves for the FP and BP probe pulses virtually
coincide and are situated closer to the $f_2$-asymptote, i.e. a
large part of energy of the third pulse is absorbed inside the
medium. However, at $\Omega_{p3}>0.8 \Omega_0$, the curves sharply
diverge: the FP probe pulse is still mainly trapped, while the BP
probe pulse is almost totally transmitted (the blue curve with
triangles gets very close to the $f_1$-asymptote in Fig.
\ref{fig3}). The energy of the first pulse is absorbed by the medium
in both cases. Thus, the most distinction between the FP and BP
cases is observed exactly in the region $\Omega_{p3} \approx
\Omega_0$. This is the asymmetry seen in Figs. \ref{fig2}(b) and
\ref{fig2}(d). At larger values of $\Omega_{p3}$, the curves
converge again, so that the medium becomes transparent for the FP
pulse as well [cf. Fig. \ref{fig2}(c)]. Finally, for the FP pulses
with $\Omega_{p3} \approx 2 \Omega_0$, the fraction of trapped
energy increases again. This allows us to see a certain periodicity
of absorption, though we will not consider the case of large
intensities ($\Omega_{p3} \geq 2 \Omega_0$) here. One can expect
more sophisticated dynamics in that case due to the process of pulse
splitting into several $2 \pi$ solitons.

The distinction between the FP and BP probe pulses is easily seen in
Fig. \ref{fig4}, which shows the peak intensity of the probe pulse
at the output versus its amplitude at the input. The BP probe pulse
behavior almost coincides with that of the single pulse (i.e., with
$\Omega_{p1}=\Omega_{p2}=0$). On the contrary, the FP probe pulse
has significantly lower value of peak intensity, which is equal to
zero at $\Omega_{p3}<1.5 \Omega_0$ (no soliton at the exit). Note
that at high input amplitudes, the secondary pulse appears as a
result of pulse splitting of both the single and BP pulses. However,
this is not the case for the FP probe pulse, i.e., the controlled
absorption blocks the splitting of a high-intensity pulse in this
case.

\begin{figure}[t!]
\includegraphics[scale=0.9, clip=]{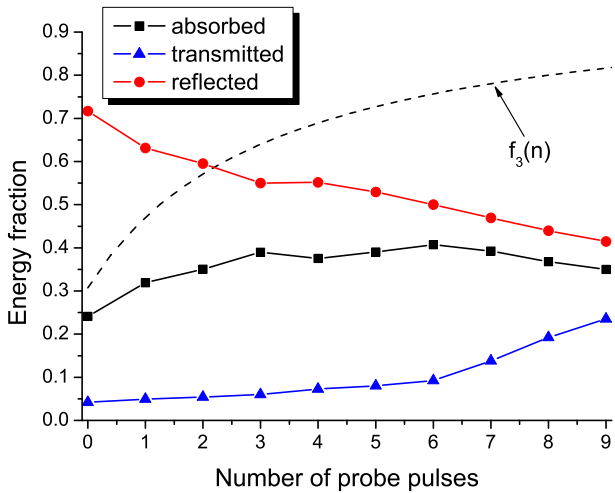}
\caption{\label{fig5} (Color online) The dependence of absorbed,
transmitted (FP), and reflected (BP) energy on the number of
additional probe pulses with the amplitude $\Omega_0$. The energy is
integrated over the time period $t=1000 t_p$.}
\end{figure}

\begin{figure}[t!]
\includegraphics[scale=0.85, clip=]{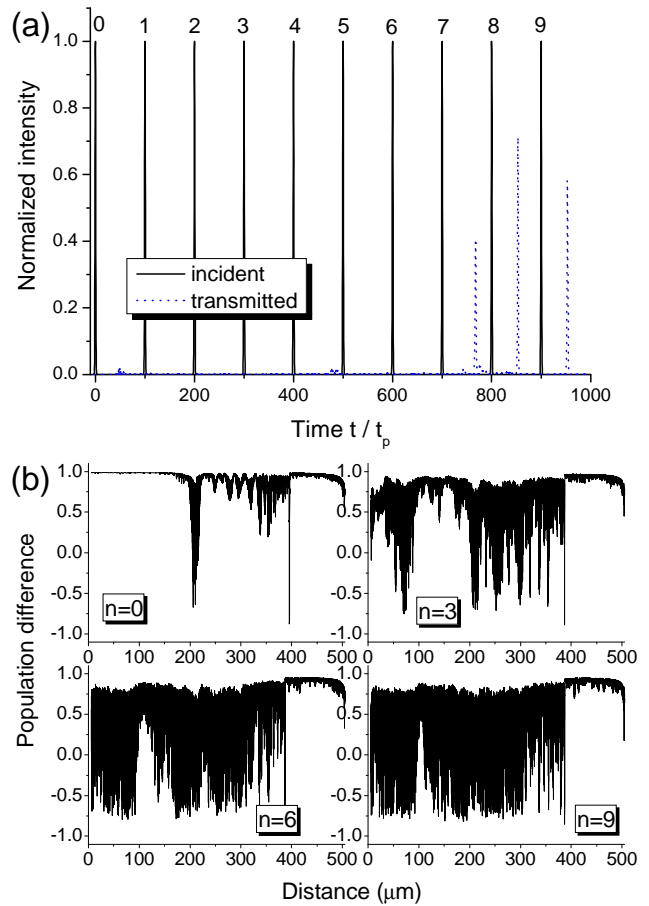}
\caption{\label{fig6} (Color online) (a) Transmitted radiation
profile for the nine incident probe pulses with the amplitude
$\Omega_0$. (b) Distributions of population difference $W$ inside
the medium at the instant $t=1000 t_p$ for different numbers of
probe pulses. Population difference $W=1$ corresponds to atoms in
the ground state.}
\end{figure}

The next question to be discussed is as follows: How many FP pulses
can be absorbed inside the medium? Obviously, the number must be
limited. To find the answer, we launch a number of identical FP
pulses (with $\Omega_{p3}=\Omega_0$) in $100 t_p$ one after another
and examine the fraction of absorbed energy as a function of probe
pulse number $n$. The asymptote for ideal trapping of all the pulses
is
$f_3(n)=(\Omega_{p1}^2+n\Omega_{p3}^2)/(\Omega_{p1}^2+\Omega_{p2}^2+n
\Omega_{p3}^2)=(1+n)/(3.25+n)$. The results of calculations for
different $n$ are presented in Fig. \ref{fig5}. It is seen that the
difference between the asymptote and the absorption curve grows
rapidly with $n$. Moreover, the absorbed energy stops increasing
already at $n \geq 3$. This means that the saturation of light
trapping takes place, while simultaneously the transmitted energy
slowly grows. The increase of transmitted energy becomes significant
at $n>6$, while the absorbed part of the energy begins to decrease.
One can expect that there should be not only the low-intensity tails
at the output, but also the solitons as well. This expectation is
proved to be valid by Fig. \ref{fig6}(a): there is no transmitted
pulse at $n=6$, but, starting from $n=7$, every extra pulse appears
at the output.

Figure \ref{fig6}(b) allows us to trace this saturation process. It
is seen that the energy of the first absorbed pulse ($n=0$) is
localized inside the medium: medium excitation near $200$ $\mu$m is
due to the collision with the BP pulse, while the postcollisional
pulse is trapped at larger distances (mainly, at $L<400$ $\mu$m). As
we launch additional (probe) FP pulses, their energy is absorbed in
the unexcited regions of the medium ($n=3$) until the distribution
of excitation becomes practically uniform ($n=6$). Note that the
region $L>400$ $\mu$m remains almost unexcited, which was
predetermined by the first pulse. This uniform distribution of
population difference corresponds to the saturation and practically
does not change when we add more pulses ($n=9$). As a result, the
interaction of the additional pulses gets weaker for $n>7$, so that
they can pass the medium. These smeared distributions also allow an
understanding of why the absorbed pulse cannot be recovered.
Therefore, we cannot talk about the storage of light and use the
term "controlled absorption". The absorbed energy is radiated by the
medium dipoles during the interval of the order of the relaxation
time.

\section{\label{time}Parameters regions for the diode action existence}

\begin{figure}[t!]
\includegraphics[scale=0.9, clip=]{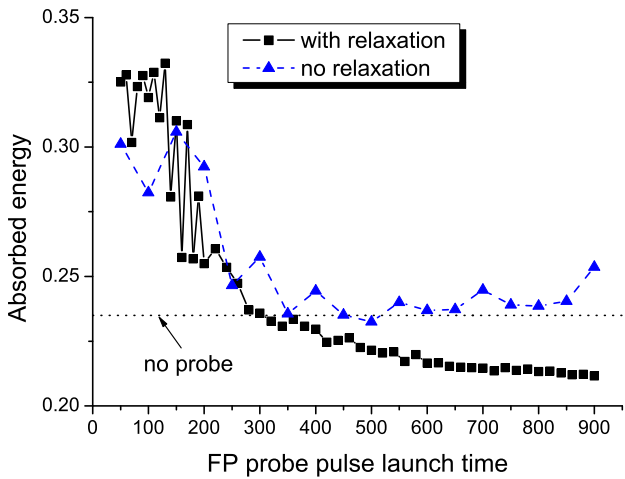}
\caption{\label{fig7} (Color online) The dependence of absorbed
energy (as a fraction of the total energy) on $t_3$, the launch time
of the probe FP pulse with amplitude $\Omega_{p3}=\Omega_0$. The
energy is integrated over the time period $t=1000 t_p$. Time is
measured in units of $t_p$.}
\end{figure}

\begin{figure}[t!]
\includegraphics[scale=0.88, clip=]{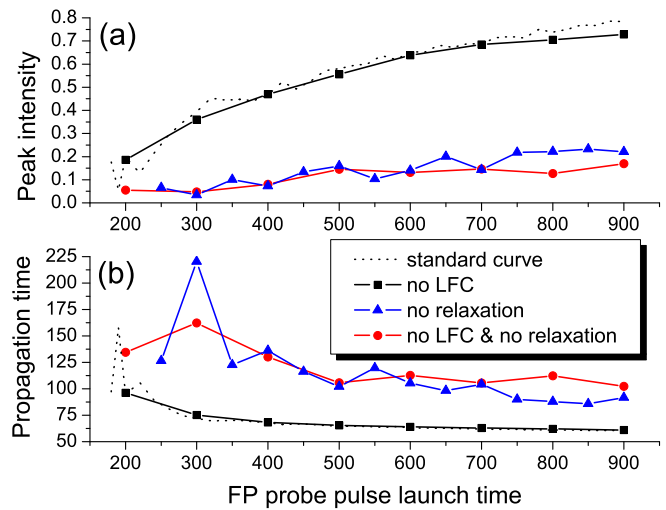}
\caption{\label{fig8} (Color online) The dependence of (a) the peak
intensity and (b) propagation time (in units of $t_p$) on $t_3$, the
launch time of the probe FP pulse with amplitude
$\Omega_{p3}=\Omega_0$. The cases of absence of local-field
correction (LFC) and relaxation are shown as well. The "standard
curve" is calculated taking into account both LFC and relaxation.}
\end{figure}

\begin{figure}[t!]
\includegraphics[scale=0.88, clip=]{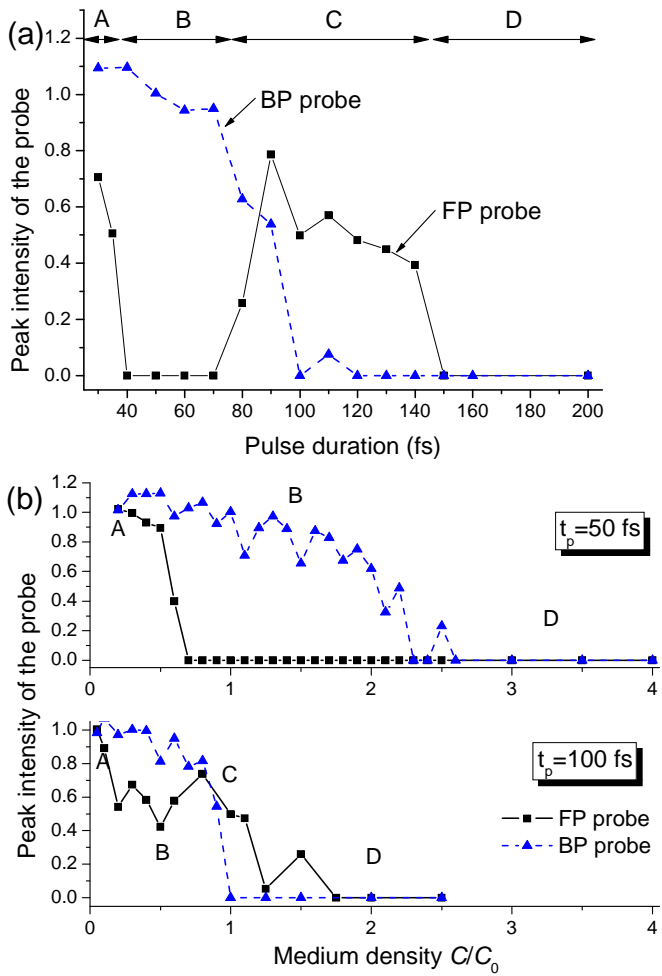}
\caption{\label{fig9} (Color online) The dependence of the peak
intensity of the probe pulse on (a) the duration of the pulses, and
(b) the density of the medium (for two different durations). The
value $C_0$ used for normalization corresponds to the Lorentz
frequency $\omega_L=10^{11}$ s$^{-1}$. In figure (a) the value
$C=C_0$ is adopted.}
\end{figure}

In the above consideration we adopted the time interval $100 t_p$
between the launch instants of the first two and the third pulses.
Here we must point out the significance of the time interval between
the pulses, which must be much less than the relaxation times to
preserve the coherence of light-matter interaction. We study this
temporal effect in Fig. \ref{fig7}, which shows the absorbed part of
the energy as a function of the launch time $t_3$ of the probe
(second FP) pulse. The first two pulses were launched at the instant
$t_1=t_2=0$. It is seen that at short $t_3<200 t_p$, absorbed energy
is somewhere between the energy of the single FP $\Omega_0$ pulse
($0.235$) and the energy of both FP $\Omega_0$ pulses ($0.47$). This
means that at the instant $t=1000 t_p$ (the final time of energy
integration), the medium contains the energy of the first FP pulse
and partly of the probe pulse. This absorbed energy slowly decreases
as we increase the launch time $t_3$, and at $t_3>300 t_p$ it drops
below the level of the single FP pulse energy. One can suggest the
main role of relaxation in this behavior. Indeed, the time, say,
$t=500 t_p=25$ ps, is comparable with the relaxation time $T_2=100$
ps adopted in our calculations. To prove this suggestion, we
performed simulations for the relaxation-free medium ($T_1=T_2=0$).
The results are presented in Fig. \ref{fig7} (the dashed line with
the triangles). It is seen that the curve in this case keeps above
the level $0.235$, i.e., the relaxation does not affect the process
of controlled absorption of the first FP pulse. However, the energy
of the probe pulse is not absorbed as well. Nevertheless, as Fig.
\ref{fig8} demonstrates, the peak intensity of the pulse transmitted
is strongly suppressed for the relaxation-free medium in comparison
with the case of the presence of relaxation (this last case is shown
by the "standard curve" in the figure). Thus, relaxation helps to
form an intense soliton during propagation through the medium
prepared by the collision of two counter-propagating pulses.

Figure \ref{fig8} also clarifies the influence of local-field
correction [the nonlinear addition to the detuning in Eq.
(\ref{dPdtau})] on the effects discussed. It is seen that this
influence can be considered as negligible for our parameters. This
is in full accordance with the conclusion of Ref. \cite{Novit1}.
Indeed, for the parameters of our calculations ($\omega_L=10^{11}$
s$^{-1}$, $t_p=50$ fs), the ratio of the pulse amplitude (peak Rabi
frequency) to the Lorentz frequency is $\Omega_0/\omega_L= \sqrt{2
\pi} \cdot 10^2 >> 1$, while, for the local-field effects to be
observed, one needs $\Omega_0/\omega_L \sim 1$.

Now let us consider the effect of pulse duration on the controlled
absorption and all-optical diode action. The results of calculations
are depicted in Fig. \ref{fig9}(a). We mark out there several
regions denoted with letters from A to D. In region A (very short
pulses, $t_p<30$ fs) both FP and BP probe pulses appear at the
output. Moreover, the total absorption of the first FP soliton is
not observed in this case. Therefore, one can say that region A is
the region of elastic collisions of the counter-propagating
solitons, which is characteristic for short enough (high-intensity)
pulses \cite{Afan, Novit2}. The region D (long pulses, $t_p>150$ fs)
is the reverse case of strongly inelastic collisions, so that the
soliton is annihilated independently of the propagation direction
\cite{Afan}. Finally, the intermediate regions B and C are of
interest for us. In region B (conditions considered above in this
paper), the FP probe pulse is absorbed, while the BP one is
transmitted. In region C the situation is reversed, so that one can
change the direction of the diode action simply by changing the
duration of the incident pulses.

Obviously, the dynamics of light depend not only on the parameters
of the pulses themselves, but also on the parameters of the medium,
such as its density. In other words, the behavior of the soliton
inside the medium is governed by the combination of the field and
matter characteristics. In the paper \cite{Shaw}, the combination
$\Upsilon \sim C t_p^2$ was proposed, so that for longer pulses, one
can obtain the effect at lower density. We study this relationship
in Fig. \ref{fig9}(b), where the probe pulse peak intensity is
plotted versus the medium density (in units of $C_0$ corresponding
to the previously used Lorentz frequency $\omega_L=10^{11}$
s$^{-1}$). One can see that at $t_p=50$ fs, the diode action covers
a wide region of densities $0.7<C/C_0<2.3$ [B region, in terms of
Fig. \ref{fig9}(a)]. At lower and higher densities, there is no
directional asymmetry in pulse behavior (elastic and strongly
inelastic conditions, respectively). If we take longer pulses
($t_p=100$ fs), region B shifts toward lower densities and becomes
narrower, in accordance with the $\Upsilon$ combination. It is also
important that the contrast between the curves in region B is not as
impressive as for shorter pulses. However, we can say with
confidence that the asymmetric propagation appears at lower
densities than previously due to the increase in pulse duration. In
addition, a narrow C region appears at $t_p=100$ fs, while region A
is located near the very vertical axis. In general, the $\Upsilon$
combination satisfactorily describes the effects of density and
pulse duration on the location and width of the region of the diode
action.

\section{\label{concl}Conclusion}

In conclusion, we studied the collisions of multiple
counter-propagating SIT solitons within the framework of the
semiclassical model of light interaction with a homogeneously
broadened two-level medium. It is shown that by means of inelastic
asymmetric collision of two proper pulses, the medium can be
prepared in such a state that it behaves as a diode transmitting the
probe pulse or not, depending on propagation direction and peak
intensity. This medium preparation is connected with the controlled
light absorption in the medium which is limited by the process of
saturation. By combining the characteristics of the medium and the
pulse, one can find the proper conditions to observe the effects of
asymmetric collisions. We should also notice that pulse retrieval
was not achieved, so that the stored energy leaves the medium
naturally as a consequence of fluorescence.

One can expect that the effects of SIT solitons collisions can be
observed in usual self-transparency media: solid dielectrics
(including ruby where SIT was first discovered), vapors of alkali
metals, molecular gases, and semiconductors. A review of the early
experiments can be found in Ref. \cite{Poluektov}. Obviously, the
solid-state systems seem to be the most convenient for our aims. The
conditions considered in this paper (very short and high-intensity
pulses, short distances of propagation) are in good agreement with
the requirements of semiconductor materials which possess short
relaxation times and comparatively high transition dipole moments.
The theoretical and experimental results on self-induced
transmission in semiconductors were reported, for example, in
\cite{Giessen, Nielsen1, Nielsen2, Smyrnov}. Another prospective
material for SIT collision experiments is the collection of
semiconductor quantum dots which can be considered as artificial
two-level atoms with high dipole moments \cite{Panzarini,
Schneider}.

The final remark deals with the problem of inhomogeneous broadening
when the resonant frequency of two-level atoms is distributed in a
certain range. Recall that our calculations were performed for the
case of a homogeneously broadened medium, i.e., for broadening only
due to the finite phenomenological relaxation $T_2$ (equal for every
atom). As a matter of fact, we are in the well-studied regime of SIT
solitons which have essentially the same main properties in both
cases of homogeneous and inhomogeneous broadening \cite{Poluektov}.
Moreover, the well-known area theorem (one of the attributes of SIT)
is not valid in a strict sense for the homogeneously broadened
medium \cite{Poluektov}, as it was confirmed recently by direct
numerical simulations \cite{Yu}. Therefore, although the role of
inhomogeneous broadening is still to be studied, we believe that the
main features of SIT pulse collisions should remain unchanged in
that case.


\begin{thebibliography}{0}
\bibitem{Gallo} K. Gallo and G. Assanto, \josab {\bf16}, 267 (1999).
\bibitem{Zhukovsky} S. V. Zhukovsky and A. G. Smirnov, \pra {\bf83}, 023818 (2011).
\bibitem{Cuicui} C. Lu, X. Hu, Y. Zhang, Z. Li, X. Xu, H. Yang, and Q. Gong, \apl {\bf99}, 051107 (2011).
\bibitem{Chunhua} C. Xue, H. Jiang, and H. Chen, Opt. Express {\bf18}, 7479 (2010).
\bibitem{Cai} X. Cai, X. Wang, and S. Li, \oc {\bf285}, 1959 (2012).
\bibitem{Feise} M. W. Feise, I. V. Shadrivov, and Y. S. Kivshar, \pre {\bf71} 037602 (2005).
\bibitem{Philip} R. Philip, M. Anija, C. S. Yelleswarapu, and D. V. G. L. N. Rao, \apl {\bf91} 141118 (2007).
\bibitem{McCall} S. L. McCall and E. L. Hahn, Phys. Rev. {\bf183}, 457 (1969).
\bibitem{Poluektov} I. A. Poluektov, Yu. M. Popov, and V. S. Roitberg, Sov. Phys. Usp. {\bf 17}, 673 (1975).
\bibitem{Novit2} D. V. Novitsky, \pra {\bf84}, 013817 (2011).
\bibitem{Bowd93} C. M. Bowden and J.P. Dowling, \pra {\bf47}, 1247 (1993).
\bibitem{Cren96} M. E. Crenshaw, \pra {\bf54}, 3559 (1996).
\bibitem{Novit1} D. V. Novitsky, \pra {\bf82}, 015802 (2010).
\bibitem{Novit} D. V. Novitsky, \pra {\bf79}, 023828 (2009).
\bibitem{Shaw} M. J. Shaw and B. W. Shore, \josab {\bf8}, 1127 (1990).
\bibitem{Afan} A. A. Afanas'ev, V. M. Volkov, V. M. Dritz, and B. A. Samson, \jmo {\bf37}, 165 (1990).
\bibitem{Giessen} H. Giessen, A. Knorr, S. Haas, S. W. Koch, S. Linden, J. Kuhl, M. Hetterich, M. Gr\"{u}n, and C.
Klingshirn, \prl {\bf81}, 4260 (1998).
\bibitem{Nielsen1} N. C. Nielsen, S. Linden, J. Kuhl, J. F\"{o}rstner, A. Knorr, S. W. Koch, and H. Giessen, \prb {\bf64}, 245202 (2001).
\bibitem{Nielsen2} N. C. Nielsen, T. H\"{o}ner zu Siederdissen, J. Kuhl, M. Schaarschmidt, J. F\"{o"}rstner, A. Knorr, and H. Giessen, \prl {\bf94}, 057406 (2005).
\bibitem{Smyrnov} O. A. Smyrnov and F. Biancalana, \prb {\bf85}, 075201 (2012).
\bibitem{Panzarini} G. Panzarini, U. Hohenester, and E. Molinari, \prb {\bf65}, 165322 (2002).
\bibitem{Schneider} S. Schneider, P. Borri, W. Langbein, U. Woggon, J. Forstner, A. Knorr, R. L. Sellin, D. Ouyang, and D. Bimberg, \apl {\bf83}, 3668 (2003).
\bibitem{Yu} X. Y. Yu, W. Liu, and C. Li, \pra {\bf84}, 033811 (2011).
\end{thebibliography}
\end{document}